\documentclass[aps,prl,twocolumn,showpacs,amsmath,amssymb,superscriptaddress]{revtex4-1}
\usepackage{graphicx}
\usepackage{bm} 
\usepackage{dcolumn}
\usepackage{amsmath}
\usepackage{amssymb}
\usepackage{multirow}
\usepackage{chapterbib} 

\begin{document}
\title{Vortex lattice melting in a boson-ladder in artificial gauge field}
\author{E. Orignac}
\affiliation{Univ Lyon, Ens de Lyon, Univ Claude Bernard, CNRS, Laboratoire de Physique, F-69342 Lyon, France}
\author{R. Citro}
  \affiliation{Dipartimento di Fisica "E.R. Caianiello", Universit\`a
  degli Studi di Salerno and Unit\`a Spin-CNR, Via Giovanni Paolo II, 132, I-84084 Fisciano
  (Sa), Italy}
\author{M. Di Dio}
\affiliation{CNR-IOM-Democritos National Simulation Centre, UDS Via Bonomea 265, I-34136, Trieste, Italy}
\author{S. De Palo}
\affiliation{CNR-IOM-Democritos National Simulation Centre, UDS Via Bonomea 265, I-34136, Trieste, Italy}
\affiliation{Dipartimento di Fisica Teorica, Universit\`a Trieste,
  Trieste, Italy}
\begin{abstract}
We consider a two-leg boson ladder in an artificial U(1) gauge field and show that, in the presence
of interleg attractive interaction, the flux induced Vortex state
can be melted by dislocations. For increasing flux, instead of the Meissner to Vortex transition in the
commensurate-incommensurate universality class, first an Ising transition from the
Meissner state to a charge density wave takes place, then, at higher flux, the melted Vortex
phase is established via a disorder point where incommensuration develops in the rung current 
correlation function and in momentum distribution.
Finally, the quasi-long range ordered Vortex phase is recovered
for sufficiently small interaction.
Our predictions for the observables, such as the spin current and the static structure
factor, could be tested in current experiments with cold atoms in bosonic ladders.
\end{abstract}
\maketitle
% \section{Introduction}\label{sec:intro}
\begin{cbunit} 
Recently, artificial gauge fields\cite{ruseckas05_gauge,*dalibard2011gauge} and artificial spin-orbit
coupling\cite{lin2011_soc,*galitski2013_soc}  have been achieved in
cold atomic gases using Raman coupling, allowing to probe the effect
of external gauge fields on interacting bosons.
The analog of the Meissner to vortex (M-to-V) phase transition for superconductors\cite{tinkham_book_superconductors} was predicted for the bosonic two-leg ladder
 in Refs.~\cite{kardar_josephson_ladder,*orignac01_meissner}.
The original proposal was made in the
context of Josephson junction ladders, where dissipation
spoils quantum coherence~\cite{fazio_josephson_junction_review}, affecting their use for superconducting qubits based circuits\cite{devoret}.
In the ultracold atomic gas a simple but already nontrivial realization
is the bosonic two-leg ladder in artificial flux\cite{atala2014},
where the M-to-V transition was observed
in non-interacting bosonic ladders at fixed flux $\pi/2$ per plaquette and
varying interleg hopping.\\
From the theoretical point of view\cite{kardar_josephson_ladder,*orignac01_meissner}, for bosons
with in-chain repulsive interactions, the M-to-V
transition falls in the commensurate-incommensurate
(C-IC) universality class
\cite{japaridze_cic_transition,*pokrovsky_talapov_prl}.
Recently it was investigated by Density Matrix Renormalization Group (DMRG) and
bosonization approach for hard-core bosons on a two-leg ladder as a
function of flux\cite{our_2015,piraud2014b} showing that the region of stability
of the M phase over the V one is largely enhanced with respect to the
non--interacting case\cite{piraud2014b}.
Moreover, besides  the
incommensuration  already predicted for
low-flux\cite{kardar_josephson_ladder,*orignac01_meissner} at  fluxes
of the order of $\pi n$, with $n$ the number of particles per rung,
a second incommensuration (2-IC) in the correlation functions is
induced by the interchain hopping\cite{our_2015,our_2016}.
However, in statistical mechanics, it is known
that  transitions in C-IC universality class can be turned into
different universality classes by various
relevant perturbations\cite{bohr1982,*bohr1982b,schulz83_cic_vortices,horowitz_renormalization_incommensurable,haldane83_cic}, thus
for a bosonic ladder in an artificial gauge field it remains a relevant question to investigate the robustness of the M-to-V phase transition.\\
In this Letter, we consider the effect of an interchain interaction ($U_\perp$)
and show that it can spoil the  M-to-V
transition, leading to the appearance of an intermediate charge density
wave phase (CDW) that can be interpreted as a melted vortex
phase. The melting of vortices is accompanied by a disorder point that
gives rise to incommensuration of correlation functions when the
density of dislocations becomes large enough to permit a greater
energy gain from the applied flux than from the 
pinning potential of the vortices.
We recall that while in two dimensions dislocations
appear at finite
temperature\cite{Berezinskii2,*kosterlitz_thouless,haldane83_cic}, in
one dimension even at zero temperature
their formation can be driven by quantum fluctuations only.\\
We consider a two-leg hard-core boson ladder in a flux $\lambda$, with Hamiltonian:
\begin{eqnarray}
H_\lambda &=&
\sum_{j,\sigma} -t \left(b^\dagger_{j,\sigma}e^{i\lambda\sigma}b_{j+1,\sigma}+
\mathrm{h. c.}\right)
\nonumber \\
&+&  \sum_j \Omega \left(b^\dagger_{j,\uparrow}b_{j,\downarrow}+\mathrm{h. c.}\right) +
U_\perp n_{j,\uparrow} n_{j,\downarrow},
\label{eq:H-model}
\end{eqnarray}
where $b_{j,\sigma}$ annihilates a boson on chain $\sigma=\pm 1/2$ at
site $j$, $n_{j,\sigma}=b^\dagger_{j,\sigma}b_{j,\sigma}$ is the
associated number operator, $t e^{i\lambda \sigma}$ the hopping amplitude along
the chains, $\Omega$  the rung hopping, and $U_\perp$  the interchain interaction.
This Hamiltonian can be mapped onto a system of spin-1/2 bosons with spin-orbit coupling
in a transverse magnetic field with each spinor state corresponding to one leg of the ladder.
In the rest of this Letter, we will consider the attractive case
($U_\perp< 0$). \\
\begin{figure}[h]
\begin{center}
\includegraphics[height=62.mm]{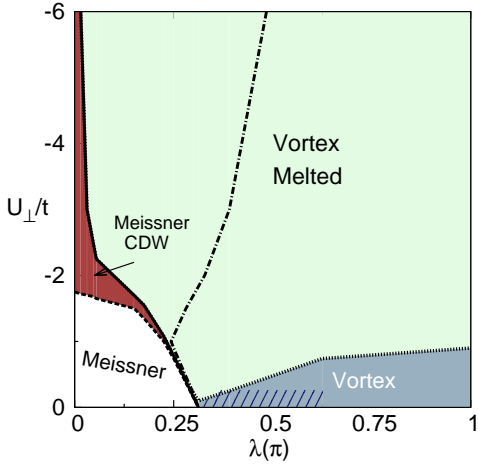}
\end{center}
\caption{Phase diagram at $n=0.5$ for a fixed value of interchain
hopping $\Omega/t=0.5$  as a function of the applied flux $\lambda$ and as
function of the strength of the interchain interaction as from DMRG simulations
for $L=64$ in PBC.
The area under the dashed black line shows the stability region
for the Meissner phase, the dark-red region between the solid and the dashed black lines shows
the region of stability for Meissner-CDW phase. The light-blue area
under the dotted line displays the stability region for the Vortex phase and shaded blue area
is the one where the second-incommensuration occurs. The light-green region above the solid and
dotted lines is where the melted or "floating" Vortex phase is stable. The dot-dashed curve
is the Lifshitz curve where the occurrence of the melted-Vortex phase is detected
in the correlation function for rung-current (see text for explanations). Color online}
\label{fig:phase_diagram}
\end{figure}
When $\Omega$ and $U_\perp=0$, the two chains are decoupled and their low energy properties are described by
the Tomonaga-Luttinger liquid model\cite{efetov_coupled_bosons,*haldane_bosons}.
For weak interchain couplings a low-energy description can be
derived\cite{kardar_josephson_ladder,*orignac01_meissner} leading to a Hamiltonian
$H_\lambda=H_c+H_s$ where:
 \begin{eqnarray}
     \label{eq:density}
     H_c&=&\int \frac{dx}{2\pi} \left[ u_c K_c (\pi \Pi_c)^2  +\frac {u_c}
    {K_c} (\partial_x \phi_c)^2 \right], \\
\label{eq:spin}
     H_s&=&\int \frac{dx}{2\pi} \left[u_s K_s \left(\pi \Pi_s +\frac{\lambda}
      {a \sqrt{2}} \right)^2 +\frac {u_s}
    {K_s} (\partial_x \phi_s)^2 \right] \\
\nonumber
   &-& 2 \Omega A_0^2 \int dx \cos
  \sqrt{2} \theta_s + \frac{U_\perp a B_1^2} 2 \int dx \cos \sqrt{8}
  \phi_s.
\end{eqnarray}
The gapless Hamiltonian $H_c$ controls the fluctuations of the
total particle density ({\it charge}) while the Hamiltonian $H_s$
controls  those of the difference
of density between the chains ({\it spin}). 
In Eq.~(\ref{eq:density})--(\ref{eq:spin}),
$A_0,B_1$ are dimensionless non-universal constants, $u_c K_c=u_s K_s = 2t
\sin(\pi n/2)$, $K_c=[1+U_\perp/(2\pi t \sin(\pi n/2))]^{-1/2}$,
$K_s=[1-U_\perp/(2\pi t \sin(\pi n/2))]^{-1/2}$ and we have assumed an
incommensurate filling of $n<1$ particles per rung.
We perform DMRG\cite{white_dmrg,*schollwock2005} simulations for this system
as a function of flux and interaction between the chains for selected fillings and values of
interchain hopping $\Omega$. We show results for $n=0.5$ and fixed interchain hopping $\Omega/t=0.5$
that are summarized in the phase diagram of Fig.~\ref{fig:phase_diagram}.
Simulations are performed for sizes up to $L=64$ in Periodic Boundary Conditions (PBC) keeping up
to $m=1256$ states during the renormalization procedure. The truncation error,
that is the weight of the discarded states, is at most of order $10^{-6}$, while the error on the
ground-state energy is of order $5\times10^{-5}$ at most.

In absence of interchain interaction and in applied flux, for moderate\cite{piraud2014b,our_2015} interchain hopping $\Omega$,
a commensurate-incommensurate (C-IC) transition\cite{japaridze_cic_transition,*pokrovsky_talapov_prl,schulz_cic2d}
occurs as a function of $\lambda$\cite{kardar_josephson_ladder,*orignac01_meissner,tokuno2014}.
In the commensurate phase, the Meissner state, the current flowing along the rungs
$J_r(l)=i\Omega (b_{l,\uparrow}^\dagger
b_{l,\downarrow}-b_{l,\downarrow}^\dagger b_{l,\uparrow})$ has
exponentially decaying correlations
$\langle J_r(l) J_r(0)\rangle \sim e^{ - |l|/\xi}$, while
the expectation value of the spin current:
\begin{equation}
J_s=-it \sum_{j, \sigma} \sigma(e^{i \lambda \sigma} b^\dagger_{j,\sigma} b_{j+1,\sigma}-
e^{-i \lambda \sigma} b^\dagger _{j+1,\sigma} b_{j,\sigma}),
\end{equation}
{\it i.e} the difference between the currents flowing along the chains, increases linearly with the flux.
The occurrence of the incommensurate phase, i.e. the establishment of
a vortex phase with quasi-long range order~(QLRO)\cite{cha2011,piraud2014b,greschner2015,greschner2016,didio2015a},
is signalled by a rapid drop in $\langle J_s\rangle$, the simultaneous appearance of two separate peaks at $k=\pm q(\lambda)/2$
in the momentum distribution
\begin{equation}
n(k)=\sum_{\sigma} n_\sigma(k)=\frac{1}{L}\sum_{\sigma}\sum^{L-1}_{i,j} e^{i k (r_i-r_j)} \langle b^+_{i,\sigma} b_{j,\sigma}\rangle
\end{equation}
and two peaks at $k=\pm q(\lambda)$ in the Fourier Transform (FT) $C(k)=\sum_l e^{-ik l}\langle
J_r(l)J_r(0)\rangle$ of the rung current correlation function.
In the vicinity of $\lambda=\pi n$, it is possible to observe, for sufficiently large $\Omega$, the
occurrence of a 2-IC\cite{our_2015,our_2016},
characterized by two satellite
peaks in $n(k)$ and $C(k)$, together with two peaks at $k=\pm \pi n$ in the spin static structure factor:
\begin{equation}
S^s(k)=\frac{1}{L}\sum_{\sigma,\sigma'}\sum^{L-1}_{j,i=0}sgn(\sigma \sigma') \langle n_{j,\sigma} n_{i,\sigma'}\rangle.
\end{equation}
When the interchain interaction $U_\perp<0$ is switched on, the last two terms in
(\ref{eq:spin}) describe the competition between the Meissner and
the in-phase CDW\cite{lecheminant2002sdsg} as a function of $U_\perp/\Omega$.
When $|U_\perp/\Omega|$ is small enough, a Meissner phase at
$\lambda=0$ is obtained, and the application of a flux is expected to
induce a C-IC transition. However,  at the C-IC transition point,
the scaling dimension\cite{schulz_cic2d,chitra_spinchains_field}
of the $\cos \sqrt{2}\theta_s$ operator is $1/(2K_s^*)=1$ implying
that  the operator
$\cos \sqrt{8}\phi_s$ of dimension $2K_s^*$ is
relevant\cite{horowitz_renormalization_incommensurable,haldane83_cic} as well.
When increasing the flux, a gapped in-phase CDW
phase\cite{mathey_pra_2009,*hu_pra_2009}  with $\langle e^{i\sqrt{2}
  \phi_s}\rangle \ne 0$ is formed instead of a vortex state with
quasi-long range order.
By duality, both the boson annihilation operators $b_{j\sigma}
\sim A_0 e^{i \frac {(\theta_c + 2 \sigma \theta_s)}{\sqrt{2}}}$ and
the rung current operator $J_r = \Omega \sin \sqrt{2} \theta_s$  present
exponentially decaying correlations, while the height of $n(k)$ at $k=0$ saturates with size because of the SRO
of $e^{i\theta_s/\sqrt{2}}$ . Since the location of the peak in $n(k)$ remains $k=0$,  this indicates that CDW retains the commensuration of the Meissner state and thus we call it Meissner-CDW (M-CDW) phase. In the statistical mechanics
context\cite{schulz83_cic_vortices,haldane83_cic},
the commensurate phase is our Meissner phase, the incommensurate phase is
the Vortex phase and the liquid phase is the M-CDW phase.
This last phase can be detected via the (in-phase) charge density
structure factor
\begin{equation}
S^c(k)=\sum_{j,\sigma,\sigma'} \langle  n_{j,\sigma}  n_{0,\sigma'}
\rangle e^{-i kj}
\end{equation}
which develops peaks at $k=\pm \pi n$ whose heights
don't scale with $L$, since the Luttinger exponent $K_c>1$ (see \cite{supplementary}).
In the lower panels $A$ and $B$ of Fig.~\ref{fig:melted_vortex_a} we follow this transition:
in the Meisnner phase (panel $A$) $S^c(k)$ is smooth and $n(k)$ shows a power-law
divergence, while in the M-CDW phase (panel $B$) $S^c(k)$ acquires the above mentioned peaks
and $n(k)$ shows a Lorentzian-like peak at $k=0$.
\begin{figure}[h]
\begin{center}
\includegraphics[height=65.mm]{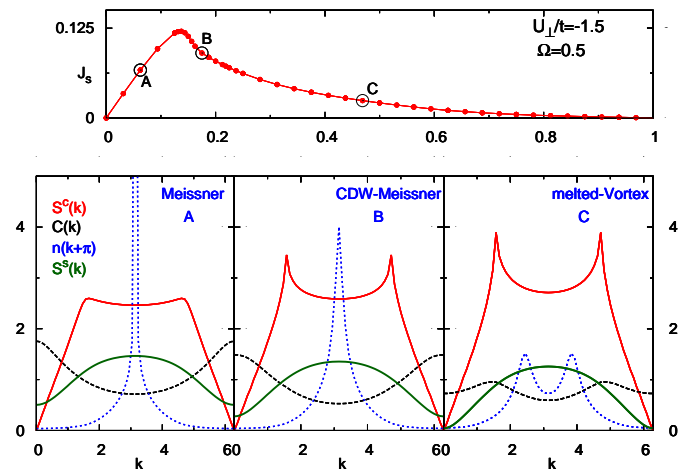}
\end{center}
\caption{ Upper panel shows the spin current $J_s$ as a function of the applied flux, red line is only a guide to eye.
Panels below show $S^c(k)$ (red solid line), $S^s(k)$ (dark-green thick solid line), $C(k)$ (black dashed line),
$n(k)$ (blue dotted line) where the argument of this last quantity has been shifted of $\pi$. Left, center and right panel
show these quantities for the cases indicated by the point $A,B$ and $C$ in the upper panel, respectively corresponding
to cases where the system is in the Meissner phase, in the M-CDW phase and in the melted Vortex phase.
Data shown are from DMRG simulations in PBC for L=64. Color online.}
\label{fig:melted_vortex_a}
\end{figure}
Meanwhile, $C(k)$ and $S^s(k)$ retain a Lorentzian shape on both sides of the transition.
For sufficiently large interchain interaction the Meissner phase is
replaced by the M-CDW even at zero flux, as shown in
Fig.~\ref{fig:phase_diagram}.
Let us note that the opening of a gap in the spin-sector prevents the observation of the 2-IC\cite{our_2015,our_2016} at increasing flux for attractive interchain interation. \\
The universality class of the flux-driven Meissner to M-CDW transition can be obtained\cite{bohr1982,*bohr1982b} by
fermionizing the Hamiltonian~(\ref{eq:spin}) in the vicinity of the
C-IC transition, as done in
Refs.\cite{wang2003field,tsvelik_field,*essler04_spin1_field,*citro02_dm_ladders}
for anisotropic spin-1 chains and spin-1/2 ladders.
The corresponding Majorana fermion Hamiltonian \cite{supplementary} has dispersion\cite{bohr1982,*bohr1982b,wang2003field}:
\begin{eqnarray}
\label{eq:eigenenergies}
E_\pm(k)^2&=&(u_s k)^2 + m^2 + h^2 + \Delta^2\\
\nonumber  &\pm& 2 \sqrt{h^2 (u_s k)^2 + h^2 m^2 +\Delta^2 m^2}.
\end{eqnarray}
where $m=-2\pi \Omega A_0^2 a$, $\Delta=-\frac \pi 2 V_\perp (B_1 a)^2$ and $h=-\frac{ \lambda u_s K_s}{a}$.
For $h_{eff}=\sqrt{h^2+\Delta^2}=|m|$, the $-$ branch of
Majorana fermions dispersion becomes gapless showing that the M-CDW quantum
phase transition falls in the Ising universality class
\cite{tsvelik_field,mccoy_revue_qft}. For $h_{eff}<|m|$
$e^{i\theta_s/\sqrt{2}}$ is long range ordered, while it is short
range ordered for $h_{eff}>|m|$. \\
At the Ising transition, the derivative of the spin current $J_s$,
 $\partial_\lambda J_s \sim  -\ln [\mathrm{max}(|\lambda-
\lambda_c|,1/L)]$  diverges logarithmically like the specific heat of the
two-dimensional Ising model\cite{mccoy_revue_qft}.
The absence of the square root threshold
singularity\cite{japaridze_cic_transition,*pokrovsky_talapov_prl} characteristic of
the C-IC transition, $\langle J_s(\lambda) -J_s(\lambda_c) \rangle
\sim \sqrt{\lambda-\lambda_c} \theta(\lambda-\lambda_c)$,
can be noted in the upper panel in
Fig.~\ref{fig:melted_vortex_a}. By plotting the numerical derivative of
$\langle J_s\rangle$ a narrow peak can be spotted at
$\lambda \simeq 0.15\pi$, past
the maximum in the spin current (see supplemental
material\cite{supplementary}).
\begin{figure}[h]
\begin{center}
\includegraphics[height=65.mm]{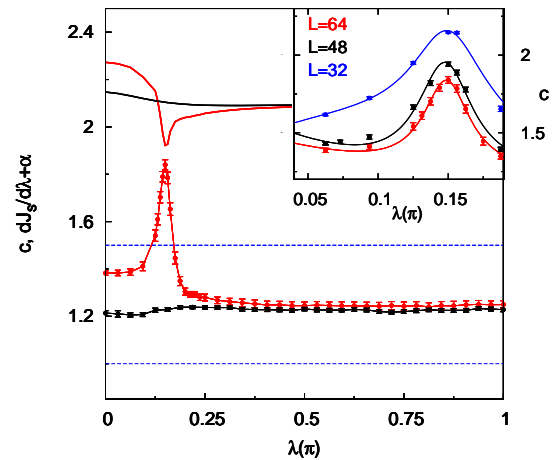}
\end{center}
\caption{In the main panel, red and black solid dots data show the central charge $c$
as a function of $\lambda$ while red and black solid lines are the numerical derivative of $J_s$
with respect to $\lambda$, respectively for $U_\perp/t=-1.5$ and $U_\perp/t=-2.0$
at $n=0.5$ and $\Omega/t=0.5$.
In the inset we show the central charge $c$ for different sizes namely $L=64,48$ and $32$, fitted with
Lorentzians curves whose spread reduces on increasing the size and whose maximum is at $\lambda=0.15 \pi$.
Data shown are extracted from Von Neumann entropy computed with DMRG in PBC
for $L=64,48$ and $32$ retaining $m=600$ states. }
\label{fig:central_charge}
\end{figure}
Another indicator of  the nature of the transition is the von Neumann entropy given by
$S_{vN}=c/3 \ln \left[ \frac{L}{\pi}\sin \left( \pi
    \frac{x}{L}\right)\right]+g$ for a system with
PBC\cite{calabrese04_entanglement} where  $c=c_c+c_s$ is the sum of the
central charges\cite{difrancesco_book_conformal} of the respectively charge and spin gapless modes and $g$ is a non-universal
constant. In the Fig.~\ref{fig:central_charge} we show the
extrapolated $c$ from fit to the numerical data of $S_{vN}$.
Despite the size effects, we can observe a  bell shaped curve centered around the critical value
$\lambda_{cdw}$ the width of which gets smaller with increasing system size. 
The height of this peak extrapolates to $c=c_c+c_s=3/2$ as size increases, indicating that the critical
point belongs to the Ising universality class
(see supplementary\cite{supplementary}), while far from $\lambda_{cdw}$, $c=c_c$ extrapolates to  unity.
Finally, the Ising nature of the transition can also be spotted looking at the value of
the peaks in the $S_c(k=\pi/2)$ that not too close to transition should be
proportional to $(\lambda-\lambda_{cdw})^{1/4}$. We have verified this behavior
for the case reported in Fig.~\ref{fig:central_charge} (see supplemental
material\cite{supplementary}).

Besides the Ising transition, the fermionized Hamiltonian also predicts\cite{schulz83_cic_vortices} the
existence of a disorder point\cite{stephenson1970a} in the crossover
to the melted-Vortex phase. Indeed, beyond a critical value
$\lambda=\lambda_d$, real space correlations function $\langle J_r(l)
J_r(0)\rangle$ and $\langle b^\dagger_{l\sigma} b_{0\sigma}\rangle$
both acquire a periodic modulation\cite{wang2003field}. Since the
wavevector of the modulation is varying with $\lambda$, the disorder
point is of the first kind.\cite{stephenson1970b}. In reciprocal
space, the modulation  gives rise to a superposition of two
Lorentzian-like peaks in $C(k)$ and $n(k)$  that are remnants of the
divergent peaks\cite{cha2011,our_2015} previously obtained in the
QLRO vortex state when $U_\perp=0$ (see panel $C$ in
Fig.\ref{fig:melted_vortex_a} and panels $A$ and $B$ in Fig.~\ref{fig:melted_vortex_b}). 
The values of $\lambda$ for which the two peaks in $C(k)$ can be resolved
are at $\lambda>\lambda_{L,C}$ with $\lambda_{L,C}$ the Lifshitz
point. One has $\lambda_{L,C}>\lambda_d$ since
resolving the peaks requires that the distance between their maxima
exceeds their width. This effect is less evident in the spin resolved $n_\sigma(k)$ where
a single Lorentzian-like peaks located at finite $k$ develops 
(black solid line in Fig.~\ref{fig:phase_diagram}).
Similar disorder and Lifshitz points have been
found in one-dimensional spin-1/2\cite{bursill1995,*deschner2013} and
spin-1\cite{schollwoeck1996,*pixley2014,*chepiga2016} chains as well
as frustrated Ising chains in transverse field\cite{beccaria2006}.
In this phase the spin and charge response functions retain respectively the 
Lorentzian shape centered around $k=0$ and the peaks at $k=\pm \pi n$.
Finally, for even higher flux and sufficiently small interaction,
$K_s^*$ becomes greater than $1$ and the operator $\cos \sqrt{8}\phi_s$ becomes irrelevant, allowing a
vortex phase\cite{kardar_josephson_ladder,*orignac01_meissner} with QLRO.
This takes place through a Berezinskii-Kosterlitz-Thouless (BKT) transition\cite{Berezinskii2,*kosterlitz_thouless} 
at the point $K_s^*(\lambda_{BKT})=1$. In Fig.~\ref{fig:melted_vortex_b} we show the recover of the QLRO vortex phase (panel $C$ from Fig.\ref{fig:melted_vortex_b}) from the melted one by decreasing
the strenght of interchain interaction $U_{\perp}/t$ for fixed applied flux $\lambda=0.625\pi$.\\

\begin{figure}[h]
\begin{center}
\includegraphics[height=65.mm]{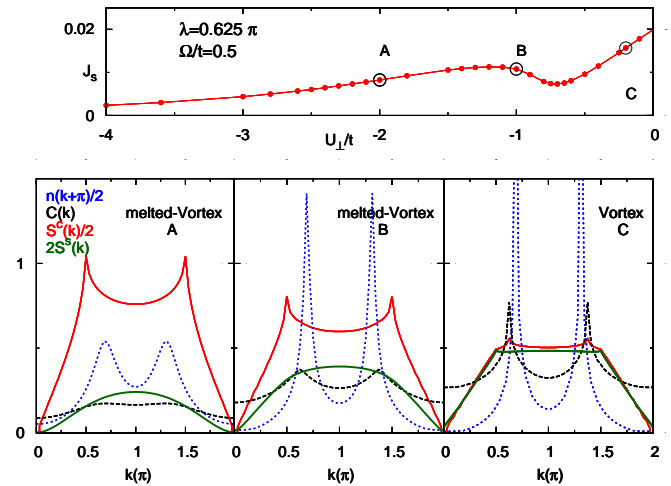}
\end{center}
\caption{ Upper panel show the spin current $J_s$ as a function of strength of interchain interaction,
solid red line is only a guide to eye. Panels below show $S^s(k)$ (dark-green thick solid line), $C(k)$ (black dashed line),
$S^c(k)$ (red solid line) and $n(k)$ (blue dotted line) where the argument of this last quantity has been shifted of $\pi$.
Left, center and right panel show these quantities for the cases indicated by the point $A,B$ and $C$ in the upper panel,
respectively corresponding to cases where the system is in the melted Vortex phase and in the Vortex phase ($C$ panel).
Data shown are from DMRG simulations in PBC for L=64. Color online.}
\label{fig:melted_vortex_b}
\end{figure}
To conclude, using bosonization and DMRG we have found (see Fig.~\ref{fig:phase_diagram}) that,
with an interchain attractive interaction, the commensurate Meissner phase and the incommensurate 
QLRO vortex phase leave space to a Meissner-CDW and to a melted vortex phase with SRO. 
Instead of having a single flux-driven M-to-V transition 
we are left with an Ising transition to the commensurate Meissner-CDW. On increasing the flux
an exponentially damped sinusoidal modulation, incommensurate with the ladder, develops in the momentum 
distribution. At the Lifshitz point $\lambda=\lambda_{L,C}$ double peaks appear in the rung current
structure factor. This indicates the existence of a disorder
point\cite{stephenson1970a,stephenson1970b} where the
bosonic Green's functions  and the rung current correlation
function develop exponentially damped oscillations in real space. The
Meissner-CDW is then crossing over into a melted vortex phase
where a SRO with proliferation of dislocations takes over.
At higher flux, a BKT transition takes place, and the
quasi-long range vortex lattice is
recovered. \\

Our predictions on the melting of vortices in Bose-Einstein condensates in optical lattices can be traced in current
experiments, where static structure factors\cite{stenger99_bragg_bec} and momentum distributions can be measured,
together with the spin current\cite{atala2014,livi}.
Using dipolar interactions, tunable by orienting the dipoles with a field\cite{kollath07_dipolar,*rydberg_atoms}, or Feshbach resonances, the interaction $U_\perp$ can be rendered attractive.\\
The detection of a melting transition as well as the non-trivial effects due to interactions can be
relevant for atomtronic ring ladders which have been proposed
for readout and gate implementation in quantum technologies\cite{Amico}, analogously to the
superconducting qubits in multi-junction circuits.

\begin{acknowledgments}
Simulations were performed at Universit\`a di Salerno, Universit\`a di Trieste and Democritos local computing facilities.
We thank M.L. Chiofalo for contributions at early stages of the work. We thank M. Capone for carefully reading the
manuscript and useful suggestions. M. Di Dio and S. De Palo thank F. Ortolani for the DMRG code and M. Dalmonte for helpful discussions.
\end{acknowledgments}
%\bibliography{revues4,habilitation,boson-soc}
%Merlin.mbs v4.21 2009-07-09.

%
\end{cbunit}

\newpage 
\widetext 

%%%%%%%%%% Merge with supplemental materials %%%%%%%%%%
%%%%%%%%%% Prefix a "S" to all equations, figures, tables and reset the counter %%%%%%%%%%
\setcounter{equation}{0}
\setcounter{figure}{0}
\setcounter{table}{0}
\makeatletter
\renewcommand{\theequation}{S\arabic{equation}}
\renewcommand{\thefigure}{S\arabic{figure}}
\renewcommand{\bibnumfmt}[1]{[S#1]}
\renewcommand{\citenumfont}[1]{S#1}

\begin{center} 
\textbf{\Large Supplementary material for ``Vortex lattice melting in a
  boson-ladder in artificial gauge field''}
\end{center}

\section{Majorana Fermions representation}
\label{sec:majoranas}
\begin{cbunit} 
Fermionization\cite{bohr1982,*bohr1982b} leads
to a  a detailed picture of the transition between the Meissner state and
the density wave states. Rescaling
$\theta_s=\sqrt{2} \theta$ and $\phi_s=\phi/\sqrt{2}$, the
Hamiltonian~([Insert Eq. number from manuscript]) is fermionized using the identities:
\begin{eqnarray}\label{eq:fermionization}
\nonumber
\frac{\cos 2 \theta}{\pi a} = -i(\psi^\dagger_R \psi_L - \psi^\dagger_L \psi_R), \quad
\frac{\cos 2 \phi}{\pi a} = i(\psi^\dagger_R \psi^\dagger_L - \psi_L \psi_R), \quad
-\frac 1 \pi \partial_x \theta = \psi^\dagger_R \psi_R + \psi^\dagger_L \psi_L,
\end{eqnarray}
yielding:
\begin{eqnarray}
  \label{eq:ham-dirac}
\nonumber
  H_s&=&-i u_s \int dx (\psi^\dagger_R \partial_x \psi_R- \psi^\dagger_L
  \partial_x \psi_L) - i m \int dx (\psi^\dagger_R\psi_L -
  \psi^\dagger_L \psi_R) - i \Delta \int dx
  (\psi^\dagger_R\psi^\dagger_L -  \psi_L \psi_R) \nonumber \\ &&
 -h
  \int dx (\psi^\dagger_R\psi_R +  \psi^\dagger_L \psi_L) + \int dx
  \frac{h^2}{2\pi u_s},
\end{eqnarray}
where:
\begin{eqnarray}
  \label{eq:ham-dirac-params}
  h= -\frac{ \lambda u_s K_s}{a}, \quad
  m=-2\pi \Omega A_0^2 a, \quad
  \Delta=-\frac \pi 2 U_\perp (B_1 a)^2.
\end{eqnarray}

Once we introduce the Majorana fermion operators
$  \zeta_{\nu,\sigma}(x)=\frac 1 {\sqrt{2}} (\psi_\nu+\sigma\psi^\dagger_\nu)(x) $
with $\nu=R,L$, $\sigma=\pm$,  the Hamiltonian~(\ref{eq:ham-dirac}) is rewritten:
\begin{eqnarray}
  \label{eq:ham-majorana}
  H_s&=&-i\frac {u_s} 2 \int dx \sum_{j=1}^2 (\zeta_{R,j} \partial_x
  \zeta_{R,j} - \zeta_{L,j} \partial_x \zeta_{L,j}) -i (m+\Delta) \int
  dx  \zeta_{R,1} \zeta_{L,1} -i (m-\Delta) \int dx  \zeta_{R,2}
  \zeta_{L,2} \nonumber \\ &&-ih    \int dx  (\zeta_{R,1} \zeta_{R,2}
  +  \zeta_{L,1} \zeta_{L,2}) + \int dx \frac{h^2}{2\pi u_s}
\end{eqnarray}
Hamiltonians of the form (\ref{eq:ham-majorana}) have previously been
studied in the context of spin-1 chains in magnetic
field\cite{tsvelik_field,*wang2003field,*essler04_spin1_field} or
spin-1/2 ladders\cite{shelton_spin_ladders,*nersesyan_biquad} with
anisotropic interactions\cite{citro02_dm_ladders}.
The quadratic Hamiltonian~(\ref{eq:ham-majorana}) can be diagonalized in
momentum space to obtain the following energy eigenvalues:
\begin{equation}
  \label{eq:eigenenergies}
  E_\pm(k)^2=(u_s k)^2 + m^2 + h^2 + \Delta^2 \pm 2 \sqrt{h^2 (u_s k)^2 + h^2 m^2 +\Delta^2 m^2}.
\end{equation}

Since observables such as the rung current are bilinear in the
Majorana fermions, obtaining their correlation functions requires
knowledge of the Matsubara Green's function for the Majorana fermions
defined as\cite{abrikosov_book}:
\begin{eqnarray}
\label{eq:green-majorana}
G_{\alpha,\beta}(x,\tau)=-\langle T_\tau \zeta_\alpha(x,\tau)  \zeta_\beta(0,0)\rangle 
 = \frac 1 {L\beta} \sum_{k,i\nu_n} G_{\alpha\beta}(q,i\nu_n) e^{i(qx-\nu_n\tau)},
\end{eqnarray}
where $\alpha=(\nu,j)$ with $\nu=R,L$ and $j=1,2$.
In Eq.(\ref{eq:green-majorana}), the Fourier space Green's function is
the $4\times 4$ matrix:
\begin{equation}
  \label{eq:green-matrix}
  G_{\alpha\beta}(q,i\nu_n)=(i\nu_n-\mathcal{H}(q))_{\alpha
    \beta}^{-1}.
\end{equation}
An explicit expression in terms of Pauli matrices is:
% Actually, some terms are missing. They come from the product of hm
% \sigma tau_2 with (z+ H(k)) This should be corrected.
\begin{eqnarray}
\nonumber
  G(k,z)&=&(z- u_s k \openone \otimes \tau_3 - h \sigma_2 \otimes \openone
  - m \openone \otimes \tau_2 - \Delta \sigma_3 \otimes \tau_2)^{-1}
  \\
  \label{eq:green-pauli}
&=& \frac{(\sigma_1 \otimes \openone)[z^2 -\mathcal{H}^2(k)](\sigma_1 \otimes \openone)[z+\mathcal{H}(k)]}{(z^2 - E_+(k)^2) (z^2 - E_-(k)^2)}
\end{eqnarray}
where $E_{\pm}(k)$ have been defined above in  (\ref{eq:eigenenergies}).\\

The real space Green's function, at equal time and zero temperature,
is defined by: 

\begin{eqnarray}
  \int_{-\infty}^{+\infty} \frac{dk }{2\pi} e^{ik
    x}\int_{-\infty}^{+\infty} \frac{d\nu}{2\pi} G(k,i\nu) =
  - \int_{-\infty}^{+\infty} \frac{dk }{2\pi} e^{ik x} 
  \left[\frac{H(k)}{2[E_+(k)+E_-(k)]} + \frac{(\sigma_1\otimes \openone)
    H(k)^2 (\sigma_1\otimes \openone) H(k)}{2E_+(k) E_-(k)
    [E_+(k)+E_-(k)]}\right]. 
\end{eqnarray}
Thus, it can be obtained from just two integrals:
\begin{eqnarray}
  \label{eq:integral-RS}
  I_1(x)&=&\int \frac{dk}{2\pi} \frac{e^{ik x}}{E_+(k) E_-(k) (E_+(k) +
    E_-(k))}, \\
 I_2(x)&=&\int \frac{dk}{2\pi} \frac{e^{ik x}}{(E_+(k) +
    E_-(k))},
\end{eqnarray}
by taking the appropriate number of derivatives with respect to $x$.

\subsection{Ising transition}
For $m=\sqrt{h^2+\Delta^2}$, $E_-(k)=u \frac \Delta m |k| + O(k^2)$ a single Majorana fermion mode becomes massless at the transition\cite{bohr1982,*bohr1982b} between the Meissner and the density wave state. This transition belongs to the
Ising\cite{mccoy_revue_qft} universality class.
\begin{figure}[h]
\centering
\includegraphics[width=9cm]{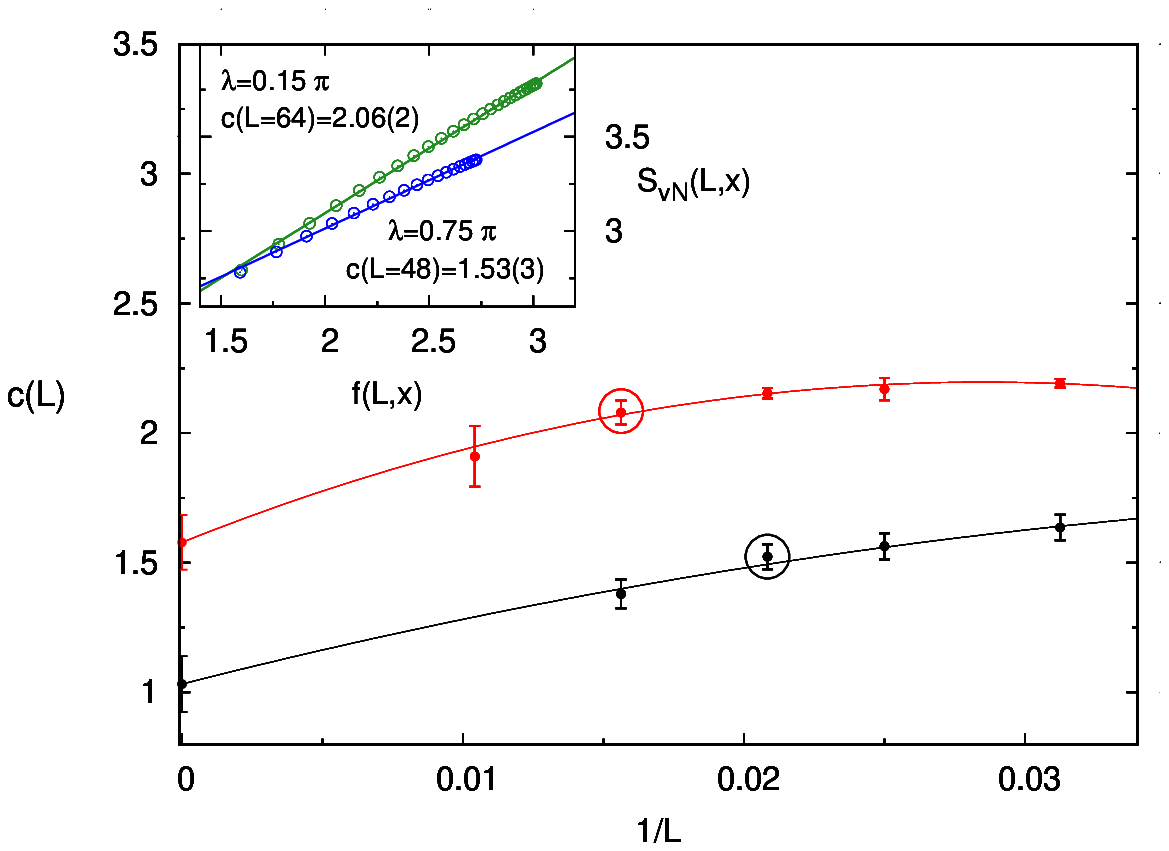}
\caption{In the main panel we show the central charge as a function of the inverse of the size $L$.
The upper curve is for $\lambda=0.15 \pi$, the critical value inferred from fitting the finite size and 
number of states $m_s$
central charges as a function of $\lambda$ (see Fig.~[Insert Fig. number from manuscript]), while the lower curve is for $\lambda=0.75 \pi$ well inside the melted Vortex phase.
The value of the central charge $c(L)$ has been extracted fitting Von Neumann entropy from simulation data
at fixed $L$ and $m_s$ as a function of $f(x)=\log(L/\pi\sin(\pi x/L))$ so that the central
charge is related to the slope of the fitting straight line. Residual dependence from $m_s$ has been removed
assuming a dependence of the form $c(m_s)=a+b/m_s^2$.
In the inset we show two typical extrapolations from raw data Von Neumann entropy at $m_s=600$.
The upper curve refers to data from PBC-DMRG simulation for the case $\lambda=0.15 \pi$ for $L=64$ while
the lower curve is for $\lambda=0.75 \pi$ at $L=48$. The $m_s$-extrapolated value for the cases plotted in the inset
panel are shown in the main panel by the red and black open dot respectively.}
\label{fig:central_charge_scaling}
\end{figure}
As a consequence, at the transition, the Von
Neumann entanglement entropy $S_{vN}=\frac 1 3 c\ln L=\frac 1 3 (c_{c}+c_{s})\ln L
= \frac 1 3 (1 + 1/2) \ln L$, where $c$ is the central charge of the gapless modes, while away from the transition it is
$S_{vN}=\frac 1 3 c_c \ln L=\frac 1 3 \ln L$ since the total density
mode $\phi_c$ is gapless. 
In Fig.~\ref{fig:central_charge_scaling} we show size extrapolation for
the central charge of two cases, at $n=0.5$, $U_\perp/t=-1.5$, $\Omega/t=0.5$, namely at $\lambda=0.15 \pi$
where we expect the Ising transition for which we get $c=1.6(1)$ and  at $\lambda=0.75 \pi$
in the melted Vortex phase where $c=1.0(1)$.
Since bosonization predicts that the
charge mode is gapless and it is described at low energy by a free boson
with $c_c=1$, the charge mode exhausts the central charge when $\lambda
\ne \lambda_{cdw}$. At $\lambda_{cdw}$, the presence of the peak indicates the appearance of an
extra gapless field with $c_s<1$, \textit{i. e.} a critical point. Conformal Field
Theories with $c_s<1$ form a discrete series\cite{difrancesco_book_conformal} with $c_s=1-\frac{6}{m(m+1)}$
with the integer $m\ge 3$. The size-extrapolated value of central charge of
the critical theory appears to be under $7/10$, leaving us with $m=3$
and $c_s=1/2$ as the only possible value, indicating that the critical
point belongs to the Ising universality class.

In Fig.~\ref{fig:central_charge_scaling} we show size extrapolation for
the central charge of two cases, at $n=0.5$, $U_\perp/t=-1.5$, $\Omega/t=0.5$, namely at $\lambda=0.15 \pi$ 
where we expect the Ising transition for which we get $c=1.6(1)$ and  at $\lambda=0.75 \pi$
in the melted Vortex phase where $c=1.0(1)$. 

\subsection{Current} 
A signature of the Ising transition can be observed also in the spin current, which is defined by
\begin{eqnarray}
\nonumber
\label{eq:meissner-current-def}
\langle J_s\rangle =\frac 1 L \left\langle \frac{\partial H}{\partial \lambda} \right\rangle =
\frac 1 L \frac{\partial E_{GS}}{\partial \lambda} =\frac 1 L \frac{\partial }{\partial \lambda}
\left( -\sum_{0<k<\Lambda} E_+(k)+E_-(k) + L \frac{h^2}{2\pi u_s}\right),
\end{eqnarray}
from which the  $\langle J_s \rangle$ in units of  $\frac
{u_s}{2a}$ is found:
\begin{eqnarray}
  \label{eq:current-integral}
 \langle J_s \rangle = \frac h {\pi u_s}   - \sum_{r=\pm 1} \int_0^\Lambda
   \frac{dk}{2 \pi} \frac{h+r\frac{h [(u_s k)^2 + m^2]}{\sqrt{h^2(u_s^2 k^2 + m^2) + m^2 \Delta^2}} }{ \sqrt{(u_s k)^2 + m^2 + \Delta^2 + h^2 +
    2 r \sqrt{h^2(u_s^2 k^2 + m^2) + m^2 \Delta^2}} }.
\end{eqnarray}
The integral (\ref{eq:current-integral}) is convergent in the limit $\Lambda \to +\infty$.

In the limit $\lambda \ll 1 $, or $h \ll 1$, 
\begin{eqnarray}
\nonumber
  \label{eq:low-flux-current}
  \langle J_s \rangle = \frac {h}{2\pi u_s} \left[3 -\frac{m^2
      -\Delta^2}{2 m \Delta} \ln \left|\frac{m-\Delta}{m+\Delta}
    \right|\right],
\end{eqnarray}
As $\Delta/m$ increases, the proportionality
constant between the flux and the current decreases but remains
positive, indicating that the Meissner effect is reduced by interchain
repulsion. 
Right at $h=m$, we can obtain the exact
expression of $\partial_h E_{GS}$ as:
\begin{eqnarray}
\nonumber
  \frac 1 L \left(\frac{\partial E_{GS}}{\partial h} \right)_{h=m} =
  \frac m {\pi u_s} \left[1-\frac \Delta m \arctan \left(\frac m \Delta
    \right)\right],
\end{eqnarray}
which shows that the current at $h=m$, \textit{i. e.}  where the
commensurate-incommensurate transition would take place in the absence of
repulsion,  is always reduced by interchain interaction.

In the vicinity of $h=\sqrt{m^2-\Delta^2}$, the dominant singularity
in $\langle J_s \rangle$ is:
\begin{eqnarray}
\nonumber
 \langle J_s \rangle^{(\mathrm{sing.})} &=& - h \left(1-\frac{m}{\sqrt{h^2+\Delta^2}}\right) \int_0^{\frac{m^2}{u_s h}}
    \frac{dk}{2\pi} \frac{1}{\sqrt{(m-\sqrt{h^2+\Delta^2})^2 +
        \frac{\Delta^2}{m^2}(u_s k)^2 }} \\
\nonumber
&& \simeq - \frac{h}{2\pi u_s} \left(\frac{\sqrt{h^2+\Delta^2} -m}
  \Delta \right) \ln \left(\frac{2 m \Delta}{h|\sqrt{h^2+\Delta^2}
    -m|}\right).
\end{eqnarray}
Thus, while $\langle J_s \rangle$ is continuous for $h \to \sqrt{m^2
  -\Delta^2}$, its derivative instead:
%$\partial_h \langle J_s \rangle $
%The derivative of the rung current is found in the form:
\begin{eqnarray}
  \label{eq:rung-curr-deriv}
  \frac{2a}{u_s} \frac{\partial \langle J_s\rangle}{\partial h} &=& \frac 1
  {\pi u_s} -\int_0^\Lambda \frac{dk}{2\pi} \sum_{r=\pm 1}
  \frac{\Delta^2}{\left[(uk)^2 + m^2 +\Delta^2 + h^2 + 2 r
      \sqrt{h^2[(uk)^2 + m^2] + m^2 \Delta^2} \right]^{\frac 3
      2}}  \\
\nonumber
&& \times \left[1 + \frac{3 m^2 [(uk)^2 + m^2]}{h^2[(uk)^2 + m^2] +
    m^2 \Delta^2} + r \frac{m^2}{\sqrt{ h^2[(uk)^2 + m^2] +
    m^2 \Delta^2}} \left(2+\frac{[(uk)^2 + m^2][(uk)^2 + m^2+\Delta^2
    + h^2]}{h^2[(uk)^2 + m^2] +
    m^2 \Delta^2}\right)\right] \nonumber .
\nonumber
\end{eqnarray}
diverges  as $ \frac{h}{4\pi \Delta a} \ln |h- h_c| $ for
$h^2=m^2+\Delta^2$.
$\langle J_s \rangle$ is a decreasing
function of $h$, and its plot as a function of $h$ presents a
vertical tangent at $h_c$. As $\langle J_s \rangle$ is increasing at small $h$ and
decreasing for $h=\sqrt{m^2-\Delta^2}$, a maximum of
the current must exist in the range
$0<h_{\mathrm{max.}}<\sqrt{m^2-\Delta^2}$ i. e. inside the Meissner phase.
%In the upper panel we can see this behaviour for Fig.~\ref{melted_n0.75}

\subsection{Charge density wave order parameter}

The CDW order is related to the $2k_F$ wave-vector component of the density operator that,
expressed using bosonization,  is $O_{CDW}(x) = A e^{i\sqrt{2} \phi_c} \cos \sqrt{2} \phi_s $
where $A$ is a non-universal constant dependent on the microscopic Hamiltonian.
On the Meissner side, $\theta_s$ is long range ordered, so the CDW correlations decay
exponentially while, on the density wave side and for $U<0$, $\langle \phi_s \rangle = 0$ and
$ O_{CDW}(x) \simeq \bar{A} \langle\cos \sqrt{2} \phi_s \rangle e^{i\sqrt{2} \phi_x} $
where $\langle \cos \sqrt{2} \phi_s \rangle \sim (a/\xi_s)^{1/8}$ with $\xi_s\sim
|\lambda-\lambda_{cdw}|$ the critical correlation length of the
Meissner-CDW transition. The exponent $1/8$ is the result of $\cos
\sqrt{2} \phi_s$ being proportional to the order parameter of the
Ising transition.   
For finite size and periodic boundary conditions the CDW correlations takes the form:
\begin{equation}
\label{eq:cdw-corr}
\langle \cos \sqrt{2} \phi_s(x) \cos \sqrt{2} \phi_s(0) \rangle 
%\langle  O_{CDW}(x) O_{CDW}(0)\rangle 
\propto \left(\frac 1
   {\xi_s} \right)^{1/4} \frac{ \cos(\pi n x) }{ \left[
 {L\sin\left(\frac {\pi x}{L}\right)}\right]^{K_c}}
\end{equation} and right at the Ising transition it becomes $ \propto 
%\langle  O_{CDW}(x) O_{CDW}(0)\rangle =B^2\cos(\pi n x)  
\left(\frac{\pi \alpha}{L \sin\left(\frac{\pi x} L \right)
    }\right)^{1/4} $   
and hence that $K_c \to K_c+1/4$ in Eq.~(\ref{eq:cdw-corr}).

Taking the Fourier transform with $2>K_c>1$ at $k=2\pi \rho_0 $ and at finite $q$
nearby, we get:
\begin{eqnarray}
  \label{eq:sc-k-general}
\nonumber
  S_c(2\pi \rho_0) - S_c(2\pi \rho_0 + q) &\propto&   \left(\frac 1
   {\xi_s} \right)^{1/4} \int_0^L \left(\frac{\pi \alpha}
 {L\sin\left(\frac {\pi x}{L}\right)}\right)^{K_c} [1-\cos (qx)] dx \\
&\propto& \frac{\pi \alpha}{\left|\cos \left(\pi \frac K 2
     \right) \right|} \left(\frac{2\pi \alpha} L \right)^{K_c-1}
 \left[\frac{\Gamma\left(\frac{K_c} 2 + \frac{L |q|}{2\pi}
     \right)}{\Gamma\left(1-\frac{K_c} 2 + \frac{L |q|}{2\pi} \right)}
   - \frac{\Gamma\left(\frac{K_c} 2
     \right)}{\Gamma\left(1-\frac{K_c} 2\right)} \right].
\end{eqnarray}
When $qL \gg 1$, one has $ S_c(2\pi \rho_0) - S_c(2\pi \rho_0 + q) \sim |q\alpha|^{K_c-1}$.
In the limit $L\to +\infty$, if $K_c>1$,  $S_c(k)$ is finite for $k \to 2\pi \rho_0$ but
presents a cusp at that point. This is consistent with the low energy predictions derived from
bosonization where $K_c=[1+U_\perp/(2\pi t \sin(\pi n/2))]^{-1/2}$.

In the transition region between the CDW and the Ising critical point, where 
the correlation length is finite but comparable to $L$, we can write:
\begin{eqnarray}
  \label{eq:ising-corr-fss}
  \langle \cos \sqrt{2} \phi_s(x) \cos \sqrt{2} \phi_s(0) \rangle =
  \left(\frac \alpha {\xi_s}\right)^{1/4} \Phi\left(\frac x L,\frac
  {L} {\xi_s} \right).
\end{eqnarray}
For given $L$ and $\xi_s$, we have to distinguish two regimes in
$q$ when we are close to the transition. For $q\xi_s \ll 1$, we recover the regime
(\ref{eq:sc-k-general}). For $q\xi_s\gg 1$, the integral in
(\ref{eq:ising-corr-fss}) becomes sensitive to the short distance
physics, and we expect: $S_c(2\pi \rho_0)-S_c(q+2\pi\rho_0) \sim
(q\alpha)^{K_c-3/4}$.
To sum up, far from the transition, we can treat the Ising order
parameter as a constant and obtain a scaling function that depends
only on $qL$ and $K_c$ 
while near the transition (with $\xi_s/L$
non-negligible) the scaling function depends on $K_c$ and both $qL$
and $q\xi_s$.

In Fig.~\ref{fig:peaks_skc} we show $S_c(2\pi \rho_0)$ as a function of the applied flux
for the case $U_{perp}/t=-1.5$ at $n=0.5$ and $\Omega/t=0.5$, far from transition 
we can detect a region on both sides of the Ising transition where this quantity is 
$\propto \xi_s ^{-1/4}$. A fit using $f(\lambda)=a/(\lambda-\lambda_{cdw})^{1/4}+b$ 
on both sides of the transition gives us as results for $\lambda_{cdw}=0.150(2)$ and $0.16(2)$,
in agreement with the estimated extracted from the central central charge or from the 
derivative of the spin current.      
\begin{figure}[h]
\centering
\includegraphics[width=9cm]{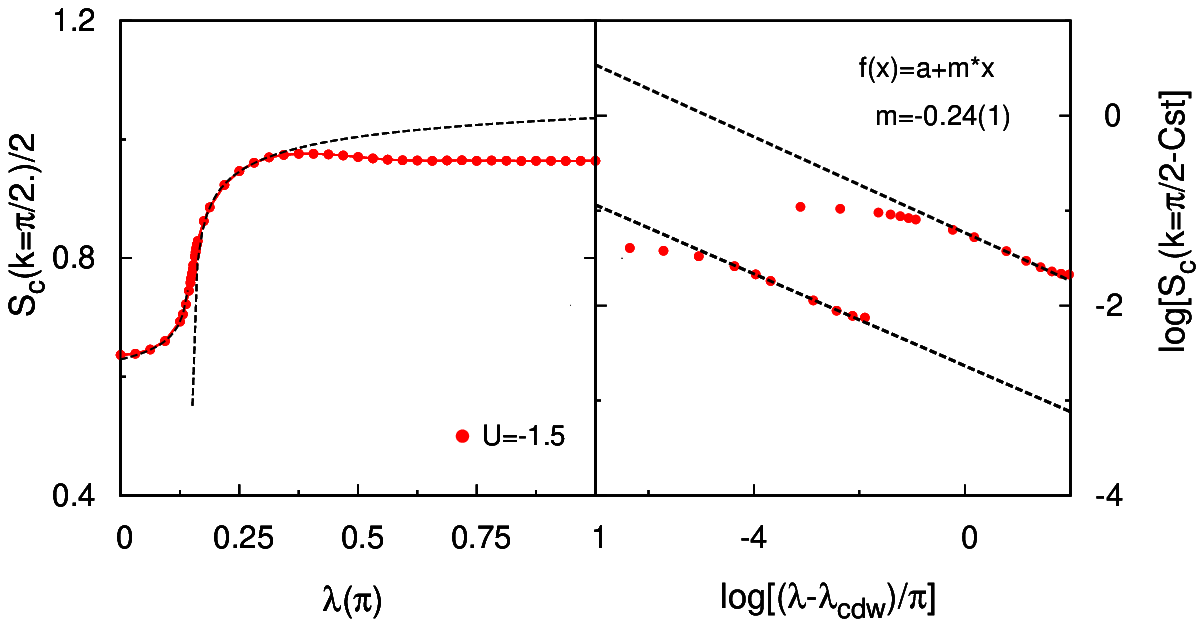}
\caption{In the left panel we show the value of the $S_c(k=\pi/2)$, {/it i.e.} where  
the correlation function develops a peak on entering the Meissner--CDW, as function 
of $\lambda$ for $U_{\perp}=-1.5$, $n=0.5$ and $\Omega=0.5/t$. 
Red dots are Data from simulation at $L=64 $ in PBC. The black dashed curves are 
fits using $f(\lambda)=a/(\lambda-\lambda_{cdw})^{1/4}+b$.
On the right panel we have plotted the $\log{S_c(k=\pi/2)}$ as a function of the logarithm of the 
deviation from the critical lambda, to find the $1/4$ exponent as the slope of the linear 
function. We have found in both cases that the slope $m=0.24(1)$ is in agreement with Ising transition.}
\label{fig:peaks_skc}
\end{figure}

\subsection{Lifshitz and disorder point}
\label{sec:dis-point}

From the spectrum of the Majorana fermion representation
(\ref{eq:eigenenergies}) we can deduce the existence of a disorder and
a Lifshitz point in some correlation functions. 
At a disorder point, a real space correlation 
function acquires a periodic modulation as a function of $x$. 
In reciprocal space, its Fourier transform is a the sum of two
Lorentzian-shaped  peaks. 
The value of $\lambda$ where the two peaks are resolved,
\textit{i. e.} where the modulation wavevector is of the order of the
peak width, is the Lifshitz point. It does not coincide with the
disorder point because of the short range order. 
This definition of disorder and Lifshitz points
applies to all correlation functions, but the existence of these
points can be inferred by considering the single particle
Green's function of the Majorana Fermions.    

The energy $E_+(k)$ is always an increasing function of $k$, while for  $2 h^2
> 2h_L^2 = m^2 +\sqrt{m^4+4m^2\Delta^2}$,  $E_-(k)$ has two degenerate
minima\cite{wang2003field} for $k=\pm k_L$ with $k_L^{(\pm)}=\pm
\frac{\sqrt{h^2 -m^2 -\Delta^2 m^2/h^2}}{u_s}$.
For $h>h_L$, $E_-(k)$ can be Taylor expanded near $k=\pm k_L$ as:
\begin{equation}
  E_-(k)^2 \simeq \Delta^2 \left(1-\frac{m^2}{h^2}\right) +
  \frac{v^4}{4 h^2} (k^2 -k_L^2)^2,
\end{equation}
and inserting in the expression of $I_1(x)$, 
  a sinusoidal modulation of the Green's function of the Majorana fermions is
expected \cite{wang2003field} at least when $h>h_L$. A more detailed
calculation would show that the sinusoidal modulation of $I_1(x)$
appears when $E_-(k)$ develops two disconnected branch cuts symmetric
with respect to the imaginary axis in complex $k$ plane. Now, the
correlation function $\langle J_\perp(x) J_\perp(0)\rangle$ 
is the trace of a product of
two Majorana Fermion Green's function and Pauli matrices, and thus
also exhibits a sinusoidal modulation of wavevector $2k_L$. So a
disorder point is present in the correlator $\langle J_\perp(x)
J_\perp(0)\rangle$, and a Lifshitz point is expected in the Fourier
transform $C(k)$. In the case of the real space Green's function of
the original bosons,
it is known that it depends on a factor coming from the charge modes
and a factor coming from Green's functions of Ising order and disorder
operators\cite{wang2003field} associated with the Majorana fermions. 
The latter factor\cite{wu_ising_correlations,wang2003field} can be expressed in terms of block Toeplitz
determinants the elements of which are Majorana fermion Green's
functions. Numerical calculations\cite{wang2003field} show that a
sinusoisal modulation appears also in the Ising order and disorder
parameter correlations. This indicates that a disorder point is also
present in the correlators $\langle b_{j\sigma}
b^\dagger_{0\sigma}\rangle$. Considering the Fourier transform, a
Lifshitz point is expected in $n(k)$. Since the disorder point is
associated with the appearance of a modulation in the Green's function
ogf the Majorana fermions, it is expected that$\langle b_{j\sigma}
b^\dagger_{0\sigma}\rangle$ and $\langle J_\perp(x) J_\perp(0)\rangle$
have a disorder point at the same value of $\lambda$. However,
in general, their Lifshitz points are not expected to coincide since  $\langle b_{j\sigma}
b^\dagger_{0\sigma}\rangle$ depends on both ``charge'' and ``spin''
modes and since the correlation lengths of the two ``spin'' parts can
differ.    
%\bibliographystyle{apsrev}
%\bibliography{revues4,habilitation,boson-soc}

\end{cbunit}

\end{document}